\documentclass[prl,aps,twocolumn,groupedaddress,floats,showpacs,final]{revtex4-1}

\usepackage{graphicx}
\usepackage{subfigure}
\usepackage{bm}
\usepackage{color}

\begin{document}

\title{Spin-Ice State of the Quantum Heisenberg Antiferromagnet on the Pyrochlore Lattice}
\author {Yuan Huang$^{1,2}$}
\author{Kun Chen$^{1,2}$}
\email{chenkun@mail.ustc.edu.cn}
\author {Youjin Deng$^{1,2}$}
\email{yjdeng@ustc.edu.cn}
\author {Nikolay Prokof'ev$^{2,3}$ }
\author {Boris Svistunov$^{2,3,4}$ }
\affiliation{$^{1}$ National Laboratory for Physical Sciences at Microscale and Department of Modern Physics, University of Science and Technology of China, Hefei, Anhui 230026, China}
\affiliation{$^{2}$Department of Physics, University of Massachusetts, Amherst, Massachusetts 01003, USA}
\affiliation{$^{3}$National Research Center ``Kurchatov Institute", 123182 Moscow, Russia }
\affiliation{$^{4}$Wilczek Quantum Center, Zhejiang University of Technology, Hangzhou 310014, China }
\date{\today}

\begin{abstract}
We study the low-temperature physics of the SU(2)-symmetric spin-1/2 Heisenberg antiferromagnet
on a pyrochlore lattice and find ``fingerprint" evidence for the thermal spin-ice state in this frustrated quantum magnet.
Our conclusions are based on the results of bold diagrammatic Monte Carlo simulations, with good convergence of the skeleton series down to the temperature $T/J=1/6$.
The identification of the spin-ice state is done through a remarkably accurate microscopic correspondence for static structure factor between the quantum Heisenberg, classical Heisenberg, and Ising models at all accessible temperatures, and the characteristic bowtie pattern with pinch points
observed at $T/J=1/6$. The dynamic structure factor at real frequencies (obtained by the analytic continuation of numerical data) is consistent with diffusive spinon dynamics at the pinch points.
\end{abstract}





\maketitle
A characteristic feature of all frustrated magnets is the close competition among
numerous classical spin configurations and the absence of an obvious
arrangement that gains the maximum amount of energy from all interaction terms~\cite{frustrated}.
Frustration prevents the development of long-range magnetic order and often
leads to novel and exotic collective phenomena. One of the best known examples
is the  spin-liquid ground state \cite{Anderson} that does not break any symmetry
and supports fractional elementary excitations and emergent gauge fields.

In many quantum antiferromagnets (AFMs), frustration has a simple geometric origin when
nearest neighbor spins form triangular or tetrahedral units.
The canonical three-dimensional example of such a system is the Heisenberg AFM
on a pyrochlore lattice that consists of corner-sharing tetrahedrons. The pyrochlore structure
is found in numerous magnetic materials and is directly associated with
such exotic low-temperature phenomena as spin glass freezing in $\textrm{Y}_2 \textrm{Mo}_2 \textrm{O}_7$ and $\textrm{Y}_2 \textrm{Mn}_2 \textrm{O}_7$ \cite{Exp_glass1, Exp_glass2, Exp_glass3}, classical spin-ice behavior in $\textrm{Dy}_2 \textrm{Ti}_2 \textrm{O}_7$ and $\textrm{Ho}_2 \textrm{Ti}_2 \textrm{O}_7$ \cite{Exp_ice1, Exp_ice2, Exp_ice3},
and cooperative paramagnetism down to ultralow temperatures in $\textrm{Tb}_2 \textrm{Ti}_2 \textrm{O}_7$ (and, presumably, a spin-liquid ground state) \cite{Exp_para1, Exp_para2, Exp_para3}.

In this Letter, we study the
SU(2)-symmetric spin-1/2 Heisenberg AFM on a pyrochlore lattice with
 \begin{equation}
 \label{hamiltonian}
	H = J \sum_{<ij>} \bm{S}_i \cdot \bm{S}_j  \qquad \qquad (J>0)\;,
\end{equation}
where $\bm{S}_i$ is the spin operator on site $i$,  and $\langle \ldots\rangle$ stands for nearest neighbor sites.
Despite its simplicity, this model is known to be notoriously difficult to solve at low, but finite,
temperature $T<J$ where perturbative treatments are not reliable; conventional Monte Carlo methods
suffer from the notorious sign problem (because of frustration); and variational
methods are not applicable. As far as we know, diagrammatic Monte Carlo (DiagMC) is the only
generic method capable of establishing controlled results in this strongly correlated regime
\cite{diagrams,generalization,triangular BDMC}, which is also the region most frequently studied experimentally.

\begin{figure}[htbp]
\centering
\includegraphics[scale=0.6, width=0.9\columnwidth]{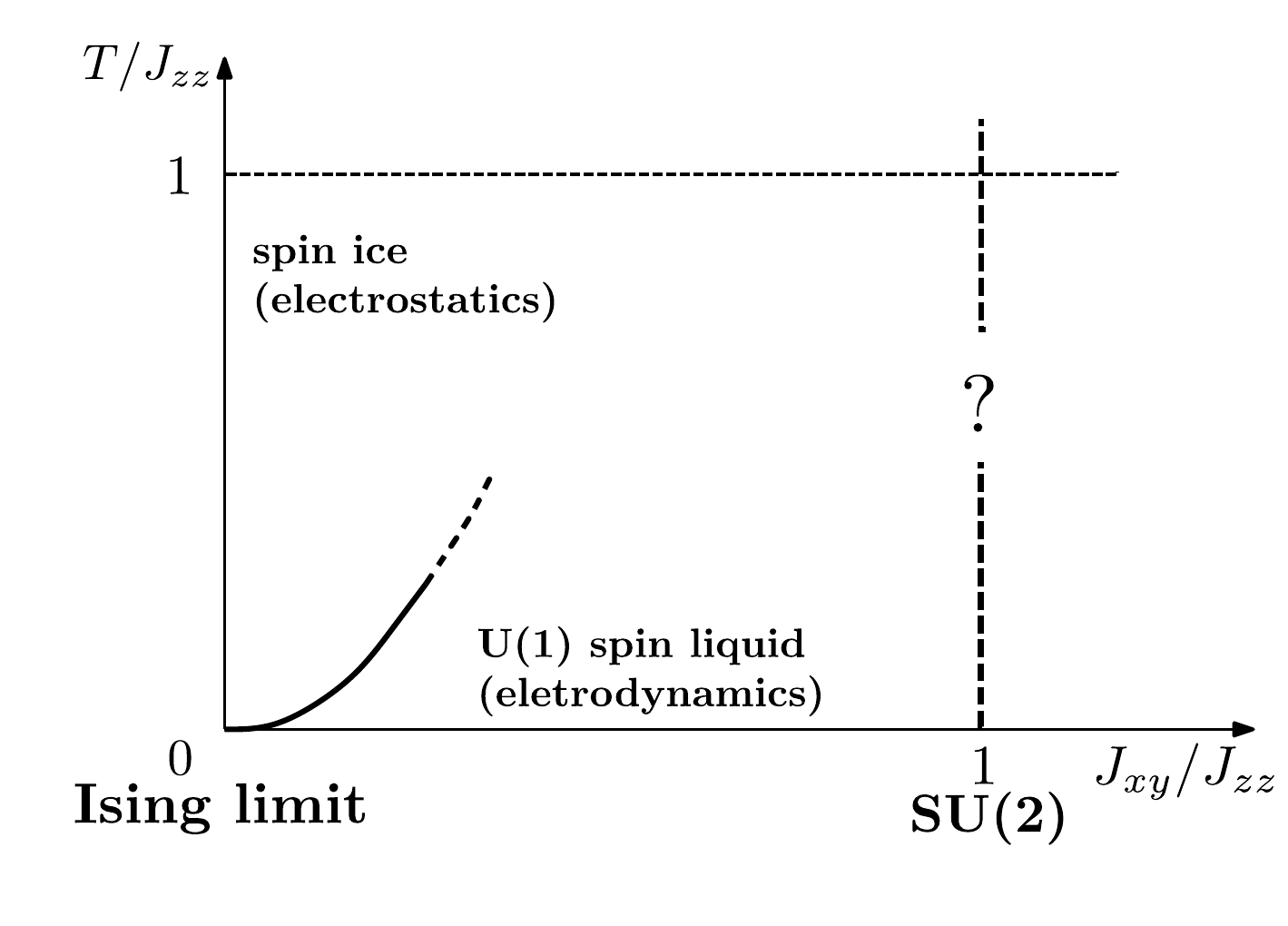}
\caption {Sketch of the finite-temperature phase diagram for the $XXZ$ model
based on perturbation theory.
For $J_{xy}\ll J_{zz}$, the first crossover at $T\sim J_{zz}$ (dotted line)
is to the thermal spin-ice state; it is followed by a second crossover at
$T \sim J_{xy}^3/J_{zz}^2$ to the low-temperature $U(1)$ spin-liquid ground state.
Whether the spin-ice state survives on approach to the isotropic Heisenberg point, $J_{xy}/J_{zz}=1$
is beyond the perturbation theory. 
}

\label{fig:xxz}
\end{figure}

Several analytic and numeric studies \cite{chalker, Canals, Fisher, Sondhi, Henley, moessner, Henley2}
looked at properties of the related $XXZ$ model
$H_{\rm XXZ}=\sum_{<ij>} J_{zz} S_i^z S_j^z + J_{xy} (S_i^x S_j^x + S_i^y S_j^y)$ that has lower
$U(1) \otimes Z_2$ symmetry, admits perturbative treatment when $J_{xy}\ll J_{zz}$, and
reduces to the Ising system at $J_{xy}=0$. At temperature $T<J_{zz}$, the Ising system
features emergent gapped $S_z=1/2$ spinons that carry fractionalized ``electric" charges
and interact by Coulomb forces; they remain deconfined because of screening. Charged excitations
``freeze out" at low temperature, leaving a massively degenerate ground state manifold.
It is known as the spin-ice phase where degenerate states satisfy the ``2-in-2-out" ice rule on each tetrahedron~\cite{anderson2} and give rise to dipolar correlations. Its characteristic feature
is the bowtie pattern with pinch point singularities in the static structure factor.
Spin-ice states were also predicted to exist in the large-$S$ and large-$N$ limits of
spin models~\cite{zinkin, chalker, Sondhi, Henley,Henley2}.


Weak transverse terms $|J_{xy}|\ll J_{zz}$ can be dealt with by degenerate perturbation theory~\cite{Fisher}.
At third order (and a low-enough temperature), quantum exchange processes
$\propto J^3_{xy}/J^2_{zz}$ operating within the hexagons are argued to lead
to the effective ``quantum electrodynamics"
type system in the continuum limit. In addition to spinons, the system features emergent gapped monopoles carrying
fractionalized ``magnetic" charges and gapless $U(1)$ gauge bosons, or ``photons"~\cite{Fisher, Bosons, jianping, QSI Rev}.
The resulting finite temperature phase diagram is illustrated in Fig.~\ref{fig:xxz}.
The ground state is argued to be a $U(1)$ quantum spin liquid with gapless ``photon" excitations.
Quantum fluctuations suppress the characteristic pinch-point singularities of the classical spin ice,
and this fact can be used for experimental identification of the spin-liquid state from the structure factor.


To answer what happens in the non-perturbative case $J_{xy}/J_{zz} \sim 1$ is a far more difficult task. In this Letter, we employ the DiagMC method to study the isotropic case $J_{xy}/J_{zz}=1$ in (\ref{hamiltonian}). We find the spin-ice state dominating system properties over a wide temperature interval, from $T \sim J$ down to the lowest simulated temperature $T=J/6$.
At $T=J/6$ the static structure factor features a characteristic bowtie pattern with pinch points.
The ultimate fingerprint evidence follows from remarkable quantum-to-classical
correspondence (QCC)~\cite{triangular BDMC} between the static spin correlation functions
of quantum Heisenberg, classical Heisenberg, and classical Ising models on the same lattice
at all length scales and all accessible temperatures. Using analytic continuation methods,
we compute the dynamic structure factor at real frequencies and observe diffusive
spinon dynamics at the pinch points and a local spin-fluctuation continuum along the nodal lines.
These results are consistent with the effective hydrodynamic theory for the spin ice~\cite{Henley,dynamic}.
A quantum spin-liquid state, if any, may emerge only at temperatures significantly below $J/6$.

{\it DiagMC and Fermionization.} The DiagMC method is a controlled numerical approach based on stochastic sampling of all
skeleton Feynman diagrams up to some high order $N$ and extrapolation to the $N\to \infty$ limit;
the series are supposed to be convergent or subject to the analytic continuation
beyond convergence radius by resummation protocols~\cite{diagrams,BDMC Unitary}.
Our implementation of DiagMC method for (\ref{hamiltonian}) is based on the $G^2W$ skeleton expansion
in the real-space--imaginary-time representation similar to that described in
Refs.~\cite{triangular BDMC}. To arrive at the diagrammatic formulation,
spins in Eq.~(\ref{hamiltonian}) are replaced with localized fermions: $\bm{S}_i = \frac{1}{2} \sum_{\alpha \beta} f^{\dagger}_{i\alpha} \bm{\sigma}_{\alpha\beta} f_{i\beta}$, where $f_{i\beta}$ is the standard
fermionic annihilation operator on site $i$, and $\bm{\sigma}$ are Pauli matrixes. Since this
procedure enlarges the Hilbert space by introducing unphysical states with zero and
double fermion occupancy, the Popov-Fedotov trick \cite{popov-fedotov,generalization} is to add a complex
chemical potential term to (\ref{hamiltonian}) to ensure exact cancellation of all unphysical contributions in
the grand-canonical statistical averages.  As a result, one ends up with the interacting
flat-band fermionic Hamiltonian
\begin{equation}
\label{PF}
	H = \frac{J}{4} \sum_{<ij> \atop \alpha \beta \gamma \delta} \bm{\sigma}_{\alpha \beta}^{\,} \cdot \bm{\sigma}_{\gamma \delta}^{\,} f^{\dagger}_{i\alpha} f_{i\beta}^{\,}f^{\dagger}_{j\gamma} f_{j\delta}^{\,}-\frac{i\pi T}{2}\sum_{i} (n_i-1),
\end{equation}
where $n_i=\sum_{\alpha} f^{\dagger}_{i\alpha} f_{i\alpha}^{\,}$.
The DiagMC method is used to sample both the auxiliary single-particle propagators
and the physical spin correlation functions. The technique allows us to go far beyond
the mean-field approximation and account for all skeleton diagrams
up to the sixth order ($>10^5$ graphs). We simulate finite systems
with periodic boundary conditions and always consider system sizes much larger
than the spin correlation length to ensure that finite-size corrections remain negligible.

{\it Correlation function.} Magnetic properties are deduced from the correlation function
$\chi(\bm{r}_i, \bm{r}_j; \tau)= \langle \bm{\hat{S}}(\bm{r}_i, 0)\cdot\bm{ \hat{S}} (\bm{r}_j, \tau) \rangle$,
where $\bm{r}_i$ is the radius vector of the lattice site $i$.
The structure factor in the momentum--Matsubara-frequency domain is given by
$S(\bm{Q},i\omega_n) = (1/V)
\sum_{i,j} \int_0^{\beta} d\tau \chi(\bm{r}_i, \bm{r}_j; \tau)e^{-i [\bm{Q} \cdot (\bm{r}_j-\bm{r}_i)+\omega_n \tau]}$
where $\bm{Q}$ belongs to the first BZ, $\omega_n=2\pi n/\beta$ is the Matsubara frequency,
and $V$ is total number of spins. The static response is described by $S(\bm{ Q},0)$, and the uniform magnetic susceptibility $\chi_u$ is given by $S(0,0)$.
\begin{figure}[htbp]
\includegraphics[scale=0.4,angle=0,width=0.95\columnwidth]{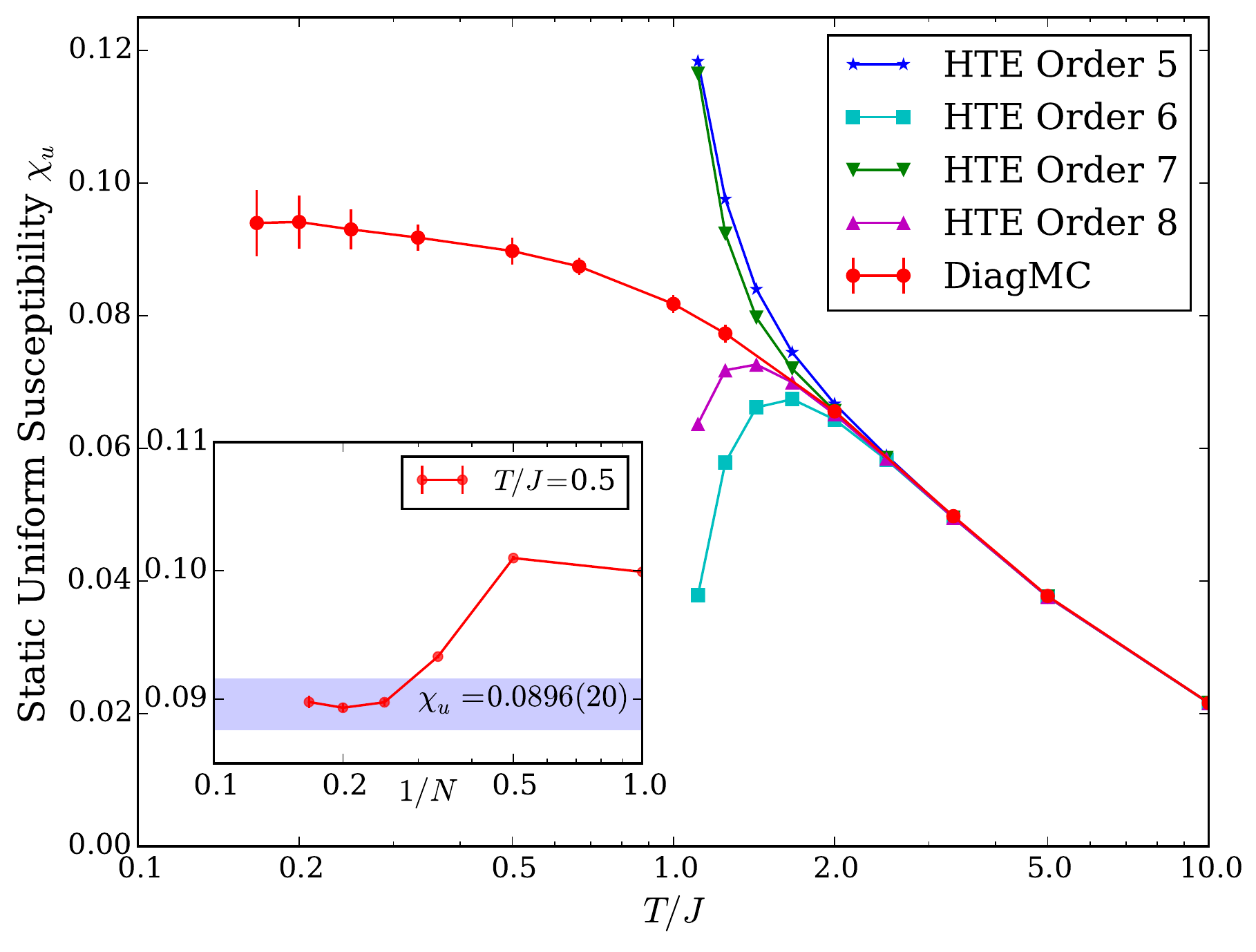}
\caption{\label{Convergence}
Uniform susceptibility $\chi_u$ as a function of temperature from the DiagMC approach(red circles)
and from the high temperature expansion (HTE) method~\cite{highT} truncated at different expansion orders.
Inset: $\chi_u$ at $T/J=1/2$ as a function of inverse maximal skeleton diagram order $N$. 
The error bar on the final answer, shown as the blue region, is a combination of statistical Monte Carlo errors for fixed-$N$ points
and the systematic error of the extrapolation to the $N\to \infty$ limit.}
\end{figure}

In Fig.~\ref{Convergence} we compare DiagMC and the high-temperature expansion~\cite{highT}
results for $\chi_u$. At high temperature $T/J>2$ the agreement between the two methods is at the
level of three meaningful digits. As temperature is lowered below $1.5 J$, the high-$T$ series explode
while the diagrammatic series continue to converge at least down to $T/J\approx 1/6$.
In the inset of Fig.~\ref{Convergence} we show how $\chi_u$ depends on the inverse diagram order
$1/N$ at $T/J=1/2$. This temperature is well below the divergence point of the
high-$T$ series and, thus, is in the strongly correlated paramagnetic regime.
Clearly, the answer does not change outside of error bars after accounting for
fifth and sixth order diagrams.
\begin{figure}[htbp]
   	\includegraphics[scale=1.0, width=0.98\columnwidth]{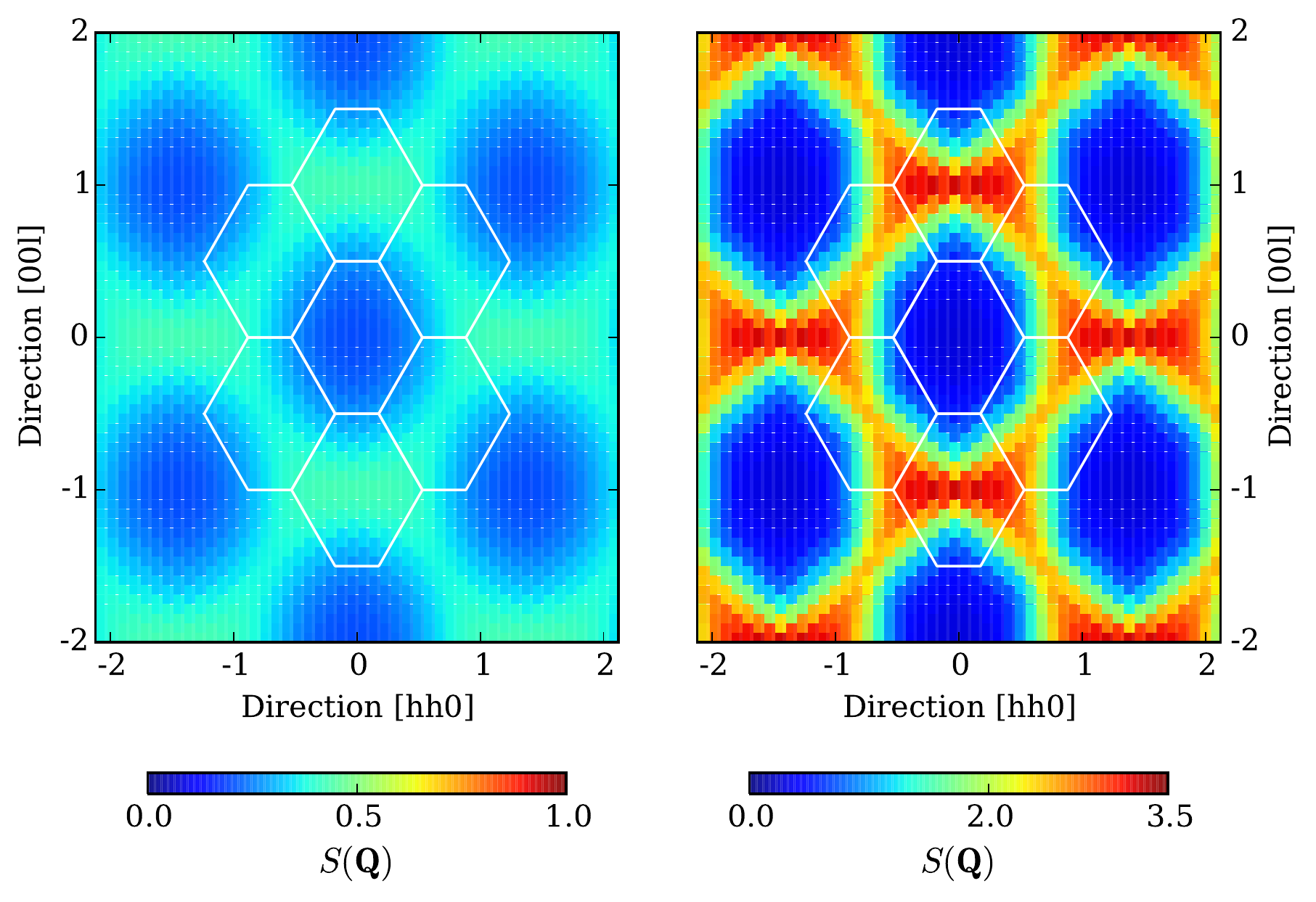}
\caption{\label{ChiK}
Structure factor $S(\bm{Q})$ in the $([hh0][00l])$ plane at $T/J=2$ (left panel) and $T/J=1/6$ (right panel).
Note that the color scheme contrast (shown at the bottom) is significantly enhanced for the left panel.}
\end{figure}

In Fig.~\ref{ChiK}, we show the evolution of the static structure factor in the $([hh0][00l])$
plane of the reciprocal space from high ($T/J = 2$) to low ($T/J=1/6$) temperature.
As the temperature is lowered, the system goes through a smooth crossover from the high-$T$
state with the checkerboard pattern in $S(\bm{ Q},0)$ to the low-$T$ state with the
bowtie pattern and pseudosingular pinch points.
As pointed out in Refs.~\cite{Henley, Sondhi}, these strongly anisotropic pinch points
are a direct consequence of the ``2-in-2-out" ice rule. All by itself, this is strong
evidence that at $T/J=1/6$ the isotropic Heisenberg model
is dominated by the spin-ice physics with excitations forming a dilute gas of electric charges.

\begin{figure}
	\centering
  		\label{fig:subfig:a} 
    	\includegraphics[scale=0.95, width=0.95\columnwidth]{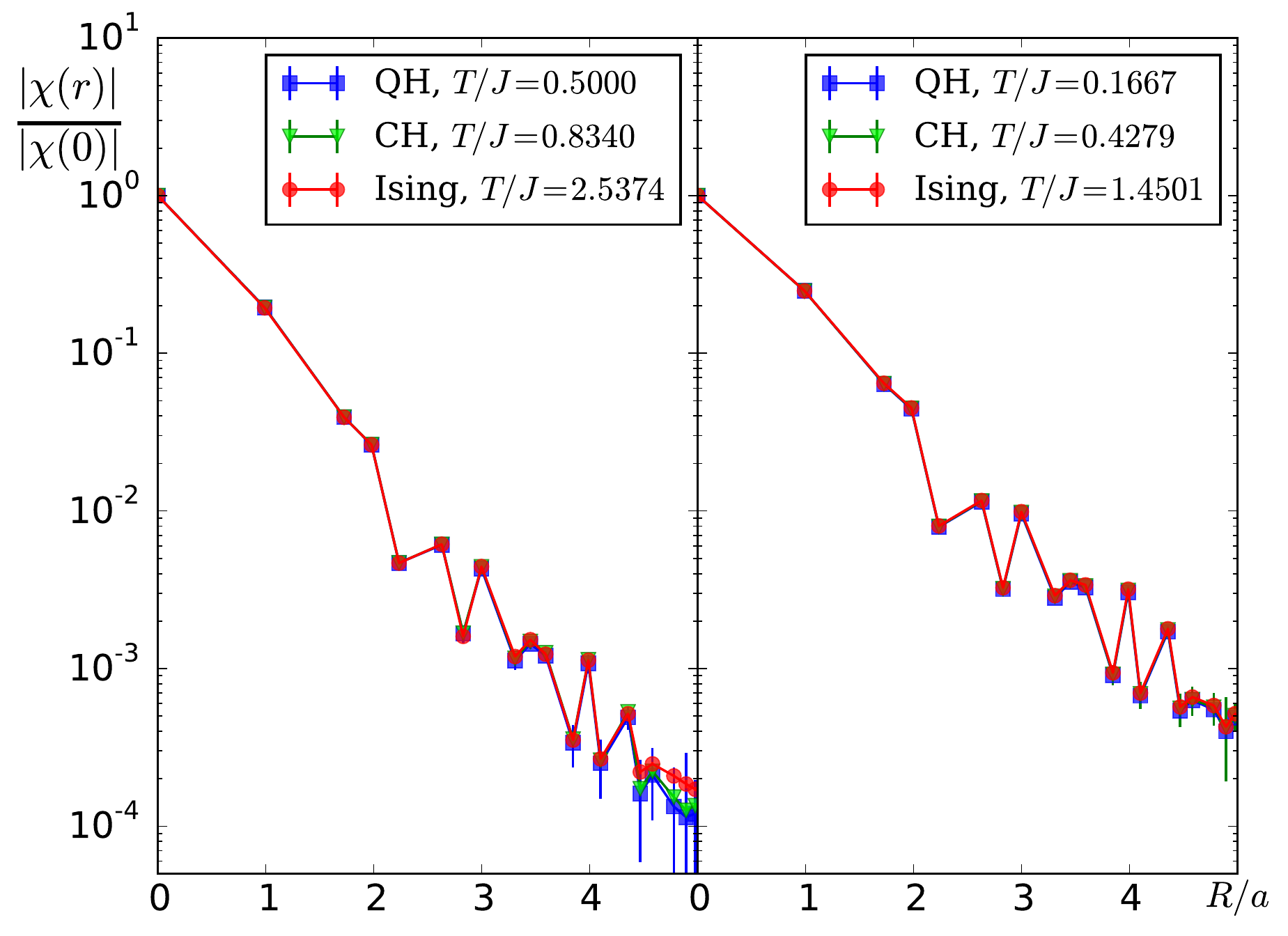}
 		\label{fig:subfig:b} 
   		\includegraphics[scale=0.95, width=0.95\columnwidth]{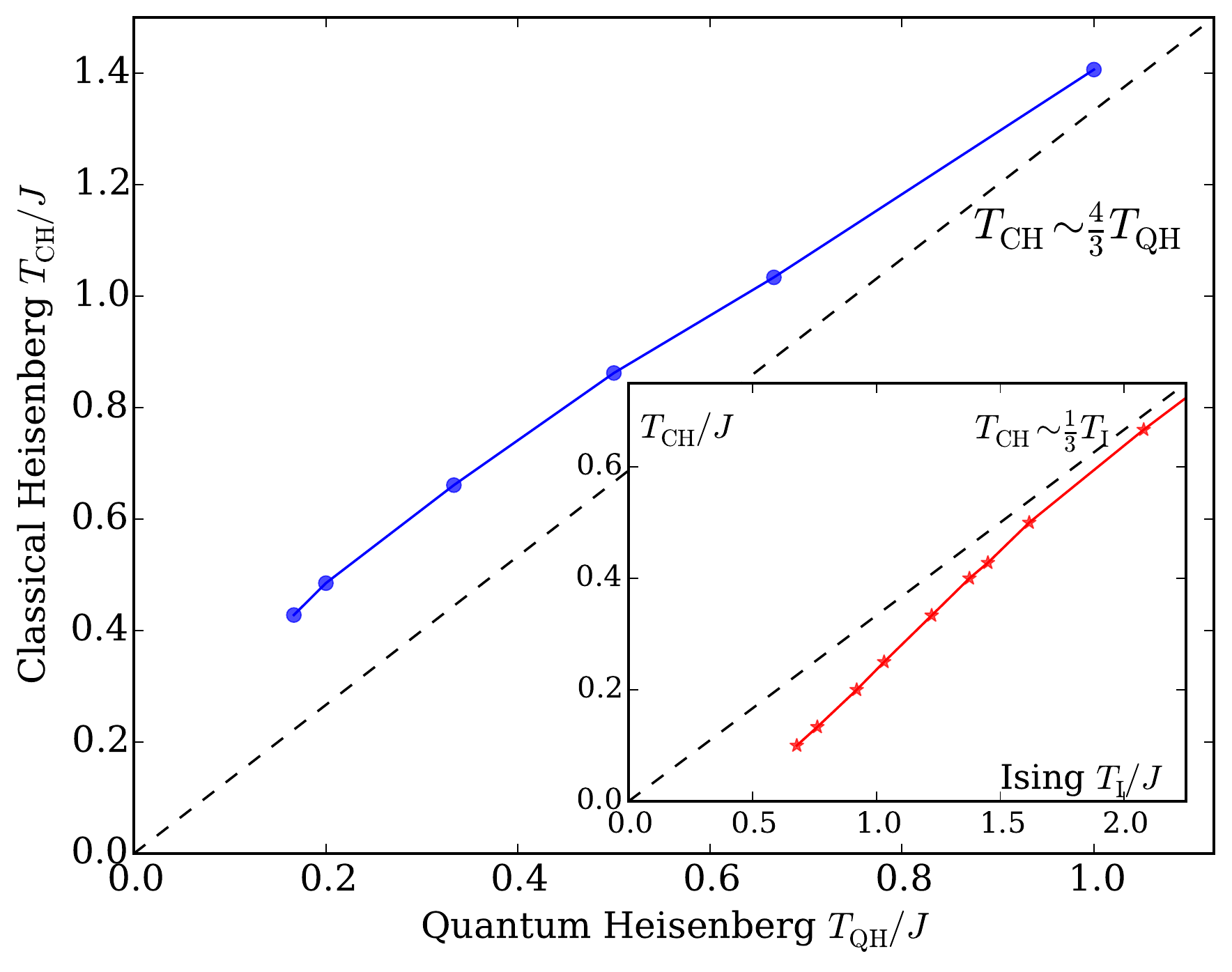}
  \caption{\label{Correspondence}
Upper panel: normalized static susceptibilities (by modulus) $|\chi_s(\bm{r}) / \chi_s({0})|$ in the quantum Heisenberg,
classical Heisenberg and classical Ising models at temperatures $T_{\rm QH}/J=1/2$, $T_{\rm CH}/J=0.8340$,
$T_{\rm I}/J=2.5374$ (left panel), and $T_{\rm QH}/J=1/6$, $T_{\rm CH}/J=0.4279$, $T_{\rm I}/J=1.4501$ (right panel).
The QCC is satisfied within the error bars at all distances.
Lower panel: quantum-to-classical temperature relationship plot $T_{\rm CH}$ {\it vs} $T_{\rm QH}$.
The straight black line is the high-$T$ relation $T_{\rm CH} = (4/3 )\, T_{\rm QH}$. Inset: temperature relationship
$T_{\rm CH}$ {\it vs} $T_{\rm I}$ between the classical Heisenberg and Ising systems.
The straight black line is the high-$T$ relation $T_{\rm CH} = (1/3) \, T_{\rm I}$.
}
\end{figure}

{\it Quantum-to-classical correspondence.}
Taking system configuration ``snapshots" is equivalent to considering multipoint
correlation functions in the diagrammatic approach (an impossible task for a large
collection of spins), not to mention that the standard technique calculates their
statistical averages. The QCC comes to rescue here.
In addition to (\ref{hamiltonian}), we consider the Ising model with spins  $s=\pm 1$ and
the classical Heisenberg model with unit-vector spins ${\bm s}$ on the pyrochlore lattice.
Both classical models have nearly identical bowtie patterns in $S({\bm Q})$
at $T=0$ \cite{Henley, Sondhi}. What we establish here, is an accurate QCC for spin correlation
functions (static in the quantum case) between the original quantum model at temperature $T_{\rm QH}$
and its classical counterparts at temperatures $T_{\rm I}$ and $T_{\rm CH}$, respectively.
The result is the fingerprint identification of dominant system configurations
at low $T$ as originating from the spin-ice state (the temperatures need to be fine-tuned
because the quantum and classical models have different spin values and configuration spaces \cite{triangular BDMC}).

The QCC protocol is as follows. For the quantum system, we compute the static correlation function $\chi(\bm r) \equiv \int_0^{\beta} d\tau \chi(\bm r_0, \bm r_i; \tau)$ where $\bm r=\bm r_i-\bm r_0$ (its classical counterparts
are defined similarly without the $\tau$-dependence). We normalize the correlation functions to unity at the origin, $f(\bm{r})=\chi(\bm{r})/\chi(0)$, and then consider the classical-model
temperature ($T_{\rm I}$ or $T_{\rm CH}$) as a free parameter to obtain the best fit for $f(\bm{r})$ curves.
The essence of QCC is that the entire functional dependence of $f(\bm{r})$ is reproduced with high accuracy
at all distances with this minimally required effort~\cite{triangular BDMC}.

Remarkably, we observe a perfect match between the quantum result at
$T_{\rm QH}$ and classical results at rescaled temperatures;
the accuracy is at the subpercent level at any temperature.
In Fig.~\ref{Correspondence}(a) we show two examples of QCC at $T_{\rm QH}/J=1/2$ and $T_{\rm QH}/J=1/6$.
Since system snapshots are readily available in the classical models, the identification of the
quantum state becomes unambiguous.
[It should be noted that the QCC is absent for the equal-time correlation function
$\chi(\bm{r},\tau =0)$.] The relationship between the temperature
of the quantum Heisenberg model and its classical counterpart is plotted in the lower panel of Fig.~\ref{Correspondence}; the relationship between the classical temperatures
is shown in the inset of the lower panel in Fig.~\ref{Correspondence}.

It is not surprising to observe the QCC in two limiting cases: (i) at high temperature $T/J \gg 1$ when weak
short-range correlations are captured at the lowest series-expansion order, and (ii) at distances much larger
than the correlation length where the statistical description in terms of the coarse-grained field becomes universal.
What we observe is different: the correspondence holds at all distances starting from the nearest-neighbor
sites and at all temperatures, including the crossover region $T/J\sim 1$. Similarly the accurate QCC
was reported for Heisenberg models on the square and triangular lattices~\cite{triangular BDMC}
(it fails in 1D). Currently, a sharp theoretical understanding of the QCC for spin-$1/2$ magnetic systems
in $D>1$ is missing.

\begin{figure}[htbp]
\includegraphics[scale=0.4,angle=0,width=0.95\columnwidth]{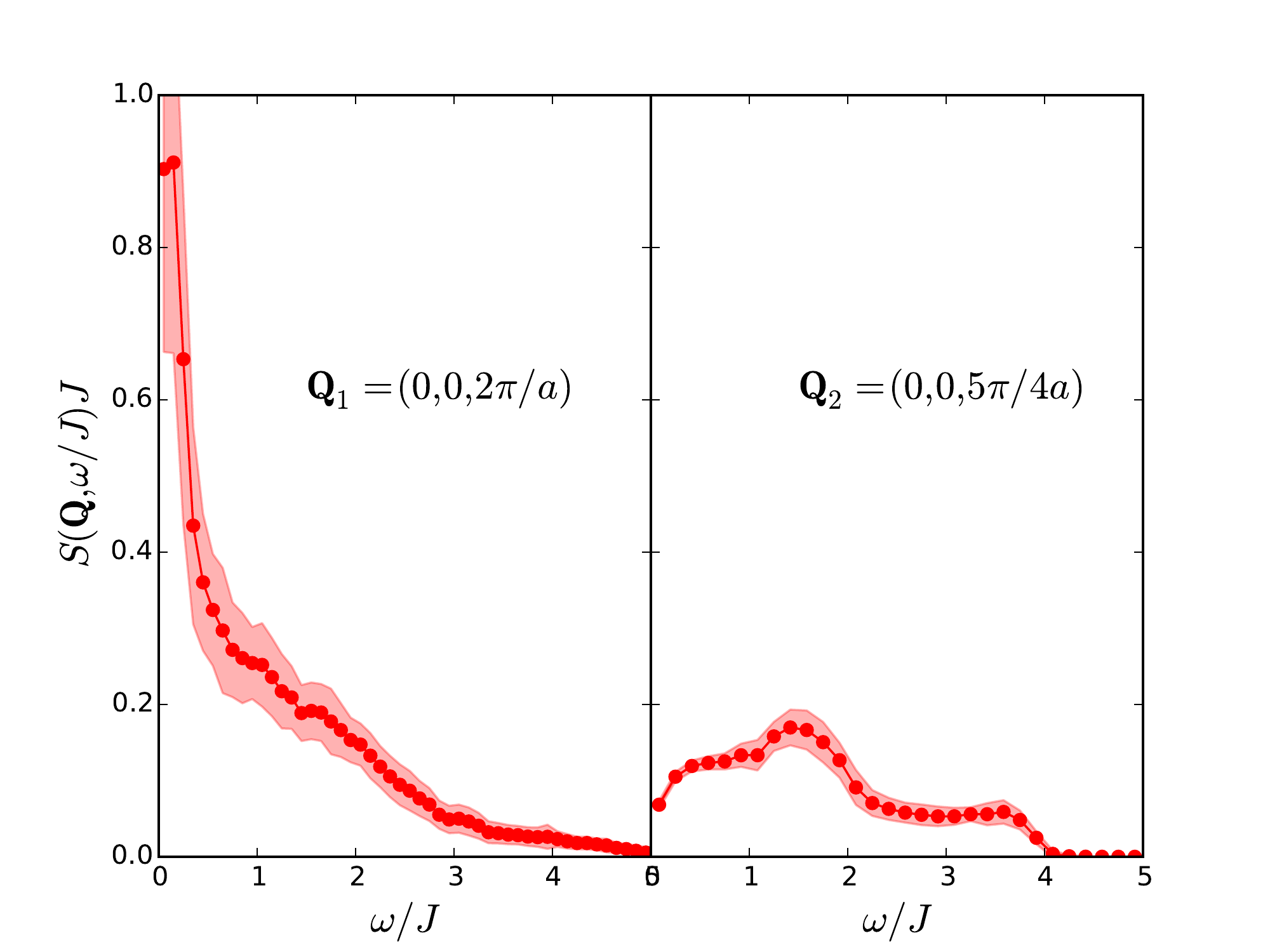}
\caption{\label{dynamic}
Dynamic structure factor as a function of frequency at the pinch point ${\bm Q}_1=(0,0,2\pi/a)$ (left panel) and on the nodal line at ${\bm Q}_2=(0,0,5\pi/4a)$ (right panel).}
\end{figure}

Having established that the static properties correspond to those of the spin ice, we proceed with the study
of the dynamic response and compute the structure factor on the real frequency axis. This quantity can be
directly measured in inelastic neutron scattering experiments. Real and Matsubara frequency functions are
related to each other by the standard linear-response theory relation
\begin{equation}
\label{analy}
S({\bm Q},i\omega_n)=\frac{1}{\pi} \int_0^{\infty} \frac{(1-e^{-\beta \omega})\omega}{\omega_n^2+\omega^2} S(\bm Q, \omega) d\omega
\end{equation}
This integral equation is solved using numerical analytic continuation methods~\cite{andrey, ana_nikolay}.
The result for two characteristic momentum points
${\bm Q}_1 =(0,0,\frac{2\pi}{a})$ and ${\bm Q}_2 =(0,0,\frac{5\pi}{4a})$,
where $a$ is the lattice constant, is shown in Fig.~\ref{dynamic}. On the basis of the thermal
spin-ice picture, we expect two dynamic contributions: one from the slow diffusive motion of spinons and the other
from propagating spin waves. At the pinch point ${\bm Q}_1$, the dynamic response is best described as that of the diffusive (Drude-type) spinon peak~\cite{Henley, dynamic}. The second point $(0,0,\frac{5\pi}{4a})$ is on one of the nodal lines, which correspond to special directions along which the spinon contribution is suppressed due to the ice rule and lattice structure~\cite{Henley}. Indeed, for this point the diffusive peak at $\omega=0$ is absent,
and a broad continuum originating from local spin fluctuations with the typical energy scale $\omega \sim J$ emerges instead.

{\it Discussion.} Using the DiagMC technique, we carried out a systematic investigation
of the quantum SU(2)-symmetric  Heisenberg AFM on the pyrochlore lattice. The correlated
paramagnetic state at temperature well below the exchange coupling constant
is unambiguously identified as the thermal spin-ice phase. The $U(1)$ spin liquid
(predicted from perturbative studies of the strongly anisotropic $XXZ$ model) has not been observed.
Apparently, the characteristic temperature to see the emergent gauge structure is much lower than $T/J=1/6$.

Our work paves the road for applications of the DiagMC technique to studies of frustrated magnetic materials
with complicated Hamilitonians when in (\ref{hamiltonian}) the exchange constant $J$ is
replaced with a $3 \times 3$ tensor and interactions are extended
beyond the nearest-neighbor sites~\cite{curnoe, Ross}. Dealing with such Hamiltonians  
does not present any additional burden for the DiagMC method
because in the skeleton formulation all lines are automatically assumed to be fully renormalized and
non-local in space-time. Our work demonstrates that it is possible to use DiagMC
to perform accurate \textit{ab initio} calculations of both static and dynamic response
for frustrated magnets, and obtain results that can be directly compared with experiments such as
the inelastic neutron scattering. In particular, one's ability to enter the strongly
correlated regime and accurately compute properties at temperatures significantly below $J$
leads to the possibility of extracting the relevant Hamiltonian parameters for frustrated
magnetic materials from measurements.

{\it Acknowledgments.}
We thank O. Starykh, G. Chen, A. Mishchenko, and especially the late C. Henley
for advice, technical assistance, and inspiring discussions. We also thank Wilczek Quantum Center for the hospitality on our visit. This work was supported by the
Simons Collaboration on the Many Electron Problem, the National Science Foundation under Grant No. PHY-1314735,
the MURI Program ``New Quantum Phases of Matter" from the AFOSR, CAS,
and the National Science Foundation of China (NSFC) under Grant No. 11275185.


\begin{thebibliography}{99}

\bibitem{frustrated}
A. P. Ramirez,
Annu. Rev. Mater. Sci. {\bf 24}, 453-480 (1994).

\bibitem{Anderson}
P. W. Anderson,
Mater. Res. Bull. {\bf 8}, 153 (1973).

\bibitem{Exp_glass1}
J. N. Reimers, J. E. Greedan, R. K. Kremer, E. Gmelin, and M. A. Subramanian,
Phys. Rev. B {\bf 43}, 3387 (1991).

\bibitem{Exp_glass2}
J. S. Gardner, B. D. Gaulin, S.H. Lee, C. Broholm, N. P. Raju, and J. E. Greedan,
Phys. Rev. Lett. {\bf 83}, 211 (1999).

\bibitem{Exp_glass3}
C. H. Booth, J. S. Gardner, G. H. Kwei, R. H. Heffner, F. Bridges, and M. A. Subramanian,
Phys. Rev. B {\bf 62}, R755 (2000).

\bibitem{Exp_ice1}
M. J. Harris, S. T. Bramwell, D. F. McMorrow, T. Zeiske, and K. W. Godfrey,
Phys. Rev. Lett. {\bf 79}, 2554 (1997).

\bibitem{Exp_ice2}
R. Siddharthan, B. S. Shastry, A. P. Ramirez, A. Hayashi, R. J. Cava, and S. Rosenkranz,
Phys. Rev. Lett. {\bf 83}, 1854 (1999).

\bibitem{Exp_ice3}
S. T. Bramwell and M. J. P. Gingras,
Science {\bf 294}, 1495 (2001).

\bibitem{Exp_para1}
J. S. Gardner, B. D. Gaulin, A. J. Berlinsky, P. Waldron, S. R. Dunsiger, N. P. Raju, and J. E. Greedan,
Phys. Rev. B {\bf 64}, 224416 (2001).

\bibitem{Exp_para2}
S. W. Han, J. S. Gardner, and C. H. Booth.
Phys. Rev. B {\bf 69}, 024416 (2004).

\bibitem{Exp_para3}
T. Fennell, M. Kenzelmann, B. Roessli, M. K. Haas, and R. J. Cava,
Phys. Rev. Lett. {\bf 109}, 017201 (2012).

\bibitem{diagrams}
N. V. Prokof'ev and B. V. Svistunov,
Phys. Rev. Lett. {\bf 99}, 250201 (2007).


\bibitem{triangular BDMC}
S. A. Kulagin, N. V. Prokof'ev, O. A. Starykh, B. V. Svistunov, and C. N. Varney,
Phys. Rev. Lett. {\bf  110}, 070601 (2013); {\it ibid.}
Phys. Rev. B {\bf 87}, 024407 (2013).

\bibitem{generalization}
N. V. Prokof'ev and B. V. Svistunov, Phys. Rev. B {\bf 84}, 073102 (2011).

\bibitem{popov-fedotov}
V. N. Popov and S. A. Fedotov, Sov. Phys. JETP {\bf 67}, 535 (1988); Proc. Steklov Inst. Math. {\bf 177}, 184 (1991).

\bibitem{chalker}
R. Moessner and J. T. Chalker,
Phys. Rev. Lett., {\bf 80}, 2929 (1998);
R. Moessner and J. T. Chalker,
Phys. Rev. B., {\bf 58}, 12049 (1998).

\bibitem{Canals}
B. Canals, and C. Lacroix,
Phys. Rev. Lett., {\bf 80}, 2933 (1998);
B.Canals, and C. Lacroix,
Phys. Rev. B, {\bf 61}, 1149 (2000).

\bibitem{Fisher}
M. Hermele, M. P. A. Fisher, and L. Balents,
Phys. Rev. B {\bf 69}, 064404 (2004).


\bibitem{Sondhi}
S. V. Isakov, K. Gregor, R. Moessner, and S. L. Sondhi,
Phys. Rev. Lett. {\bf 93}, 167204 (2004).

\bibitem{Henley}
C. L. Henley,
Phys. Rev. B {\bf 71}, 014424 (2005).

\bibitem{moessner}
R. Moessner, and A. P. Ramirez,
Phys. Today, {\bf 59(2)}, 24 (2006).

\bibitem{Henley2}
C. L. Henley,
Annu. Rev. Condens. Matter Phys. {\bf 10}.1146 (2010).

\bibitem{anderson2}
Anderson PW, Phys. Rev. {\bf 102}, 1008 (1956)

\bibitem{zinkin}
M. P. Zinkin, M. J. Harris, T. Zeiske,
Phys. Rev. B {\bf 56}, 11786 (1997).

\bibitem{QSI Rev}
M. J. P. Gingras and P. A. McClarty,
Rep. Prog. Phys. {\bf 77}, 056501 (2014).

\bibitem{Bosons}
A. Banerjee, S. V. Isakov, K. Damle, and Y. B. Kim,
Phys. Rev. Lett. {\bf 100}, 047208 (2008).

\bibitem{jianping}
J. P. Lv, G. Chen, Y. Deng, and Z. Y. Meng
Phys. Rev. Lett. {\bf 115}, 037202 (2015).

\bibitem{dynamic}
P. H. Conlon and J. T. Chalker, Phys. Rev. Lett. 102, 237206 (2009).

\bibitem{BDMC Unitary}
K. Van Houcke, F. Werner, E. Kozik, N. V. Prokof'ev, B. V. Svistunov, M. J. H. Ku, A. T. Sommer, L. W. Cheuk, A. Schirotzek,  and M. W. Zwierlein,
Nat. Phys. {\bf 8}, 366370 (2012).



\bibitem{highT} H. J. Schmidt, A. Lohmann, and J. Richter,  Phys. Rev. {\bf B 84}, 104443 (2011).

\bibitem{andrey}
A. S. Mishchenko, N. V. Prokof'ev, A. Sakamoto, and B. V. Svistunov,
Phys. Rev. B {\bf 62}, 6317 (2000);
A. S. Mishchenko, in \emph{Correlated Electrons: From Models to Materials},
eds. E. Pavarini, W. Koch, F. Anders, and M. Jarrell, Forschungszentrum Julich GmbH, (2012).

\bibitem{ana_nikolay}
N. V. Prokof'ev and B. V. Svistunov, JETP Lett. {\bf 97}, 649 (2013).

\bibitem{Ross}
K. A. Ross, L. Savary, B. D. Gaulin, and L. Balents
Phys. Rev. X {\bf 1}, 021002 (2011).

\bibitem{curnoe}
S. H. Curnoe,
Phys. Rev. B {\bf 78}, 094418 (2008).

\end{thebibliography}
\end{document}